\documentclass[showpacs,twocolumn]{revtex4}
\usepackage{amsmath,amsfonts,amssymb}
\usepackage[dvips]{graphicx}
\def\lm {\lambda}
\def\lmeff {\lambda_{\rm eff}}

\def\rs {\rho_s}

\begin{document} 
\title{An exactly solvable model of reversible adsorption on a disordered substrate}
\author{J. Talbot$^1$, G. Tarjus$^2$ and P. Viot$^2$}
\affiliation{$^1$Department   of  Chemistry  and  Biochemistry,
 Duquesne University,Pittsburgh, PA 15282-1530\\ $^2$Laboratoire de
 Physique Th{\'e}orique de la Mati\`ere Condens\'ee, Universit{\'e} Pierre et
 Marie Curie, 4, place Jussieu,75252 Paris Cedex, 05 France}

\begin{abstract}
We consider the reversible adsorption  of dimers on a regular lattice,
where   adsorption occurs on    a finite fraction   of sites  selected
randomly.  By comparing this system to the pure system where all sites
are  available for adsorption, we show  that when the activity goes to
infinity, there  exists a   mapping between this  model and   the pure
system at the same density. By  examining the susceptibilities,  we
demonstrate that there is no mapping at finite activity. However, when
the  site density  is small  or moderate,   this mapping  exists up to
second order in site density. We also propose and evaluate approximate
approaches that may be applied to systems where no analytic result is known. 
\end{abstract}

\maketitle

\section{Introduction}

Over the years a significant and sustained research effort has been
directed at understanding the effect of disorder in adsorption
processes\cite{RE92,D01}. This is commensurate with the widespread
occurrence of disorder in many different kinds of adsorption, e.g.
gases on solid surfaces and in porous media \cite{KRTM96},
biomolecules \cite{JWA96} and colloidal particles \cite{AWM02,ASWM02}
adsorbing from solution onto solid surfaces, and catalysis
\cite{OBB03,Oshanin2003}.

From  a theoretical perspective, the  challenge is  how to incorporate
disorder into statistical  mechanical descriptions  of the equilibrium
and  kinetic  properties.   In the absence of exact results, 
it is desirable to develop approximate treatments, possibly
by mapping the    system containing disorder to a pure system 
with no disorder.

One  way  to represent disorder is  the  random site  surface (RSS) in
which adsorption  sites  are uniformly and  randomly  distributed on a
plane. The adsorbent  molecules, represented by  hard spheres, bind to
these immobile sites. Steric exclusion is incorporated in the model in
that a site  is available for adsorption  only if the nearest occupied
site  is   at  least one   particle  diameter   away.  The   model  is
characterized by the dimensionless site density.

This model was originally studied in the context of irreversible
adsorption where it was shown that there exists a mapping to an
irreversible adsorption process on a continuous surface
\cite{JWTT93}. The existence of this mapping, which is exact and valid in 
any dimension, means that from a knowledge of the amount adsorbed as a
function of  time  in a continuous  space we  can  calculate the amount
adsorbed  on  the  RSS surface     of  given  site   density  at   any
time.  Adamczyk and coworkers  have successfully applied the model and
its extensions (e.g. allowing for finite size adsorption sites) to the
adsorption of latex spheres on mica surfaces\cite{AWM02,ASWM02}. 

More  recently, Oleyar and Talbot   \cite{OT07} studied the reversible
version of the RSS model in which hard spheres  adsorb and desorb from
immobilized sites in a plane. Here the quantities  of interest are the
adsorption  isotherm, i.e. the  amount adsorbed  as  a function of the
bulk phase  activity of the solute  and the structure of  the adsorbed
layer.    Somewhat  surprisingly,   the   theoretical  description  of
the system in equilibrium (even when no phase transitions intervene)   
is more difficult than   in the irreversible case
since  there  appears   to  be  no  exact  mapping  to  the reversible
adsorption on a  continuous  surface. Moreover, in one-dimension  where
the exact  solution   of the  corresponding   model without  disorder,
i.e. hard rods  on a line is well  known \cite{T36}, attempts to solve
the equilibrium RSS have so far  proved unsuccessful (despite the fact
that  many one-dimensional  statistical  mechanical models  have exact
solutions).

It   is the purpose  of this  article to  study  a  simpler model, the
adsorption of   dimers  on a  one-dimensional   lattice.   Disorder is
introduced by randomly eliminating a given fraction  of the sites. The
advantage of the lattice model is that essentially exact solutions are
available for both the pure and disordered systems. This permits us to
examine  the existence of a possible  mapping between the two. We show
that there  is  a mapping  in  the  limit  of small and   large
activities, but not in general. In the absence of a full mapping it is
still useful  to   investigate approximate approaches that  permit  an
accurate description    of  the  thermodynamics  of  the    disordered
system. We note that Oshanin
{\it et al}.\cite{OBB03,Oshanin2003} studied a similar lattice 
model for the catalytic reaction $A+A\to0$. 

We show that the introduction of an effective activity leads to highly
accurate estimates of the  thermodynamic properties of the  disordered
system.    This methodology   can  be generalized   to more realistic,
off-lattice models.

\subsection{The pure model}

The system  consists of adsorbed  molecules  which are  in equilibrium
with  a bulk  phase containing an adsorbate  at  activity $\lambda$.  The
molecules bind to sites of a one-dimensional lattice. Occupancy of one
site excludes occupancy of the nearest neighbor sites: See Fig.
\ref{fig:purelattice}. We note  that  this model is isomorphic  to the
adsorption of dimers  on the dual  lattice  \cite{E93}.  For  a system
consisting of $N$ adsorption sites  the adsorbed phase can be formally
described with the grand canonical partition function:

\begin{figure}[t]

\resizebox{8cm}{!}{\includegraphics{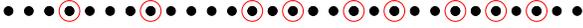}}
\caption{Adsorption on a pure lattice. The filled circles represent the adsorption sites ($N = 30$),
while the open circles show the adsorbed molecules ($n = 9$). Steric exclusion prevents the 
simultaneous occupancy of any two adjacent sites.}\label{fig:purelattice}
\end{figure}

\begin{equation}\label{eq:1}
\Xi^*(\lambda,N)=\sum_{\{z_i=0,1\}}\prod_{i=1}^{N-1}\lambda^{z_i}(1-z_iz_{i+1})\lambda^{z_N}
\end{equation}
here  $z_i$ is the site    occupancy, $\lambda=\exp(\beta\mu)$ is  the
activity  and  free   boundary  conditions  are  imposed.   We use the
superscript $^*$ to denote the  pure system, i.e. one in the 
absence of disorder.  The exact calculation  of
the partition function is a simple exercise using the transfer matrix
approach: See the appendix.

The result can be expressed as
\begin{align}\label{eq:pfp}
\Xi^*(N,\lambda)=1+\sum_{n=1}^{n_{\rm max}}\left(\begin{array}{c}
	N-n+1\\
	n
	\end{array}
\right)\lambda^n
\end{align}
where $n_{\rm max}=[(N+1)/2]$ is the maximum number of dimers that can be
adsorbed on the lattice of size $N$ ($[x]$ represents the integer part of $x$)\cite{OBB03,Oshanin2003} .

The average number of dimers adsorbed on a chain of $N$ sites at an 
activity $\lambda$ is given by 
\begin{equation}\label{eq:2}
{\cal N}_N(\lambda)=\lambda\left(\frac{ \partial \ln \Xi^*(\lambda,N) }{\partial\lambda }\right)
\end{equation}
and the fraction of occupied sites can be computed as
\begin{equation}
\rho^*(N,\lambda)=\frac{{\cal N}_N(\lm)}{N}
\end{equation}
which in the thermodynamic limit is given by
\begin{equation}\label{eq:13}
\rho^*(\infty,\lambda)=\frac{2\lm}{(1+\sqrt{1+4\lm})\sqrt{1+4\lm}}
\end{equation}
This has the expected behavior in the limits of small (Langmuir isotherm)
and large (half of the sites are occupied) activities.
The susceptibility, or fluctuation in the number of adsorbed molecules, is given by
\begin{align}
\chi (\rho )=\frac{<n^2>-<n>^2}{N}=\lambda \frac{\partial\rho}{\partial\lambda} 
\end{align}
after some calculation one obtains that, in the thermodynamic limit,
\begin{align}\label{eq:10}
\chi^*  (\rho  )=\rho (1-\rho)(1-2\rho)
\end{align}

\subsection{Model with disorder}
\begin{figure}

\resizebox{8cm}{!}{\includegraphics{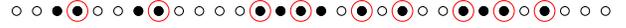}}
\caption{Adsorption on a diluted lattice. Out of a total of 30 sites 14 (filled) 
are available for adsorption. These sites form clusters 
of lengths 2,2,4,1,1,3, and 1}\label{fig:lattice}
\end{figure}

We now consider  a  diluted site model   in which only a  fraction  of
randomly selected sites  are present. The  grand partition function in
this case is given by
\begin{equation}
 \Xi(\lambda,\{ \eta\} )=\sum_{\{ \eta
\} }\sum_{z_i=0,1}\prod_{i=1}^{N-1}\lambda^{z_i\eta_i}(1-z_iz_{i+1}\eta_i\eta_{i+1})\lambda^{z_N\eta_N}
\end{equation}
where $\eta_i=0,1$ denotes the absence or presence  of an adsorption  site $i$.
The probability of finding an adsorbing site is given by
\begin{equation}
P(\eta)=\rho_s\delta_{\eta,1}+(1-\rho_s)\delta_{\eta,0}.
\end{equation}
There are no correlations between sites. A sample configuration is
shown in Fig. \ref{fig:lattice}.

For a given site configuration $\{ \eta_i\}$, the number of clusters of exactly $l$ contiguous adsorption sites is given by
\begin{equation}
n_l(\{ \eta_i\})=\sum_{i=1}^{N}(1-\eta_{i-1})\eta_i\ldots\eta_{i+l}(1-\eta_{i+l+1}).
\end{equation}
The mean number of clusters of size $l$ is then given by
\begin{equation}\label{eq:3}
\overline{n_l}=N(1-\rho_s)^2\rho_s^l.
\end{equation}
One can check that the sum rule characterizing  the total number of occupied sites is verified, namely
\begin{equation}\label{eq:4}
\sum_{l=1}^\infty l\overline{n_l}=N\rho_s.
\end{equation}
Note that the thermodynamic limit has been taken in the two equations:
corrections  for a finite  system occurs  for $1-\rho_s\simeq 1/N$ and
need not be included in these calculations.

Since  the adsorption  site  occupancies are  quenched variables,  the
average over disorder is taken over the logarithm  of the partition 
function, with the following result:
\begin{equation}
\overline{\ln(\Xi(\rho_s, \lambda))}=\sum_{l=1}^{\infty}\overline{n_l}\ln\Xi^*(\lambda,l),
\end{equation}
where  we have used the  result for the  partition function of a finite
lattice of $l$  connected sites  (i.e. the pure model) 
with free  boundary  conditions.  The mean
density of adsorbed dimers is given by
\begin{align}\label{eq:11}
\rho(\rho_s,\lambda)&=\lambda \left(\frac{ \partial  \overline{\ln(\Xi(\rho_s, \lambda))}}{\partial\lambda }\right)
&=\sum_{l=1}^{\infty}\overline{n_l}{\cal N}_l(\lambda)
\end{align}
where 
\begin{equation}
{\cal N}_l(\lambda)=\lambda\frac{\partial\ln(\Xi^*(l,\lm))}{\partial\lambda}
\end{equation}
is the average number of dimers on a (full) lattice of size $l$ with free boundary conditions.

For small site density, one easily obtains
\begin{align}\label{eq:rsmall}
&\rho(\rho_s,\lambda)={\frac {\lambda}{1+\lambda}}{\rho_s}-2\,{\frac {{\lambda}^{2}}{
 \left( 1+\lambda \right)  \left( 1+2\,\lambda \right) }}{{\rho_s}}^
{2}\nonumber\\
&+{\frac { \left( 3+2\,\lambda \right) {\lambda}^{3}}{ \left( 1+2\,
\lambda \right)  \left( 1+3\,\lambda+{\lambda}^{2} \right)  \left( 1+
\lambda \right) }}{{\rho_s}}^{3}\nonumber\\
&-2\,{\frac {{\lambda}^{4} \left( 3\,
\lambda+2 \right) }{ \left( 1+4\,\lambda+3\,{\lambda}^{2} \right) 
 \left( 1+3\,\lambda+{\lambda}^{2} \right)  \left( 1+2\,\lambda
 \right) }}{{\rho_s}}^{4}+ O \left( {{\rho_s}}^{5} \right) 
\end{align}

On the other hand, whatever the site density, 
when the activity $\lambda\to  \infty$, ${\cal N}_l(\lambda)\to   p$ for $l=2p$ and
${\cal N}_l(\lambda)\to p+1$  for $l=2p+1$, so that one obtains
\begin{equation}
\rho(\rho_s,\infty)=(1-\rho_s)^2\sum_{p=1}^\infty p\rho_s^{2p}\left(1+\frac{1}{\rho_s}\right),
\end{equation}
which gives
\begin{equation}\label{eq:9}
\rho(\rho_s,\infty)=\frac{\rho_s}{\rho_s+1},
\end{equation}
a result identical to the case of hard rods onto a line with random sites\cite{TTV07}.
(Note, however, that $\rs$ cannot take any positive real value and is bounded by 1).

In order to investigate the existence of a mapping between this model and
the  pure model,     we consider the   thermodynamic  quantities,   in
particular we focus on the susceptibility.

In the presence  of disorder, there are  two kinds  of susceptibility:
the  connected susceptibility $\chi_c(\rho_s,\lambda)$ describing  the thermal
fluctuations of the density of adsorbed particles 
and    the   disconnected  susceptibility     $\chi_d(\rho_s,\lambda)$
describing  the disorder-induced fluctuations  of the density of   adsorbed particles.

The connected susceptibility is given by the thermodynamic relation \cite{RST94}
\begin{equation}
\chi_c(\rho_s,\lambda)=\lambda\frac{\partial\rho(\rho_s,\lambda)}{\partial\lambda}.
\end{equation}
For   large activities, the  connected  susceptibility  goes to $0$ as
$1/\lambda$. This result has a  simple physical interpretation. For  a
given  realization of the disorder,   the  densest   configuration is   selected  when
$\lambda\to  \infty$. In this  limit  fluctuations are suppressed, and
the connected susceptibility vanishes.

The disconnected susceptibility is given by the second cumulant (in the average over the disorder) 
of the number of adsorbed dimers for a given configuration of the  disorder, $\sum_{l}n_l(\{ \eta_i\}){\cal N}_l(\lambda)$:
\begin{equation}\label{eq:6}
\chi_d(\rho_s,\lambda)= \frac{1}{N}\sum_{l,l'=1}^\infty (\overline{n_ln_{l'}}-\overline{n_l}\,\overline{n_{l'}}){\cal N}_l(\lambda){\cal N}_{l'}(\lambda)
\end{equation}

The different terms of the disorder average $\overline{n_ln_{l'}}$ can
be obtained   by  sorting out  configurations  of  overlapping    and
non-overlapping clusters:
\begin{enumerate}
\item When $l=l'$, there is a contribution when the two clusters are in the same location, and $\overline{n_ln_{l'}}=\delta_{ll'}N(1-\rho_s)^2\rho_s^l$.
\item When two clusters have a boundary in common, $ \overline{n_ln_{l'}}=2N(1-\rho_s)^3\rho_s^{l+l'}$, the factor $2$ coming from the two possibilities, right and left. 
\item When two clusters have overlapping sites, $\overline{n_ln_{l'}}=0$ and the number of possibilities is equal to $(l+l'+1-\delta_{ll'}$. 
\item For all other  configurations, there is no overlap between clusters and $\overline{n_ln_{l'}}=\overline{n_l}\,\overline{n_{l'}}$: this corresponds to $N(N-l-l'-3)$ configurations. 
\end{enumerate}
After some calculation, one obtains that
\begin{align}\label{eq:5}
&\overline{n_ln_{l'}}-\overline{n_l}\,\overline{n_{l'}}=N(1-\rho_s)^2\rho_s^l\nonumber\\
&\left(\delta_{ll'}+2(1-\rho_s)\rho_s^{l'+1}-(l+l'+1)(1-\rho_s)^2\rho_s^{l'}\right).
\end{align}

In the large-activity limit,
$\lambda\to \infty$, one easily shows from Eqs. (\ref{eq:pfp}) and (\ref{eq:2}) that 
\begin{equation}\label{eq:7}
\sum_{l=1}^{\infty}\rho_s{\cal N}_l(\infty)=\frac{\rho_s}{(1-\rho_s)^2(1+\rho_s)}
\end{equation}
and 

\begin{equation}\label{eq:8}
\sum_{l=1}^{\infty}l\rho_s{\cal N}_l(\infty)=\frac{\rho_s(1+\rho_s+2\rho_s^2)}{(1-\rho_s)^3(1+\rho_s)^2}
\end{equation}
From Eqs~(\ref{eq:6}), (\ref{eq:5}), (\ref{eq:7}) and (\ref{eq:8}) the following expression of $\chi_d(\rho_s,\infty )$ now results:
\begin{equation}
\chi_d(\rho_s,\infty )=\frac{\rho_s(1-\rho_s)}{(1-\rho_s)^3}.
\end{equation}
Finally, by inverting Eq.(\ref{eq:9}), the disconnected susceptibility is obtained as
\begin{equation}
\chi_d(\rho_s,\infty )=\rho (1-\rho)(1-2\rho)
\end{equation}
which  is {\em identical to the susceptibility in a pure  system at the same
density} (Eq.(\ref{eq:10})).  Since $\chi_c(\rho_s,\infty)=0$, this result establishes that the 
{\em total} fluctuations of the adsorbed number of dimers is the same in the disordered
system at infinite activity and in the pure system at the same mean adsorbed density, $\rho$. 
This points to the existence of an exact mapping between the two systems under these
conditions. The mapping can actually be proven by extending the above considerations 
to the correlation functions. 

Is the mapping exact for finite activities? In order to answer this question, we 
compare $\chi(\rho(\rho_s,\lambda))$ and $\chi_c(\rho_s,\lambda)+\chi_d(\rho_s,\lambda)$ for small values of $\rs$. The first
orders of the expansions read
\begin{align}
& \chi^*(\rho(\rho_s,\lambda))= (\frac {\lambda } {\lambda +1}\rho_s\nonumber\\
&-\frac {{\lambda}^{2} \left( 8\lambda+5 \right) }{ \left( 1+2\lambda  \right)  \left( \lambda +1
 \right)^{2}}{\rho_s}^{2}\nonumber\\
&+{\frac {{\lambda}^{3} \left( 18\,{
\lambda }^{3}+69\,{\lambda}^{2}+66\,\lambda+17 \right) }{ \left( 1+2\,
\lambda  \right)  \left( 1+3\,\lambda+{\lambda}^{2} \right)  \left( 
\lambda +1 \right) ^{3}}}{\rho_s}^{3}, \nonumber\\
&+O \left( {{\rho_s}}^{4} \right) )
\end{align}
where we have combined Eqs. \ref{eq:10} and \ref{eq:rsmall},
and 
\begin{align}
&\chi_c(\rho_s,\lambda)+\chi_d(\rho_s,\lambda)=
\frac {\lambda}{\lambda+1}\rho_s\nonumber\\
&-{\frac {{\lambda}^{2} \left( 8
\,\lambda+5 \right) }{ \left( 1+2\,\lambda \right)  \left( \lambda+1
 \right) ^{2}}}{\rho_s}^{2}\nonumber\\
&+{\frac {{\lambda}^{3} \left( 18\,{
\lambda}^{2}+49\,\lambda+17 \right) }{ \left( 1+3\,\lambda+{\lambda}^{
2} \right)  \left( \lambda+1 \right) ^{2} \left( 1+2\,\lambda \right) 
}}{\rho_s}^{3}\nonumber\\
&+O \left( {\rho_s}^{4} \right) ).
\end{align}

It is noticeable that the expansions coincide at first and second order, 
but differ at third order and above.  We speculate
that this property is  general in the  sense that it is independent of
the model   (dimers, k-mers, hard  rods,...) and  of the dimension. In
addition, one can see that the two  expansions approach the same limit
when the activity is very large. This means that they differ maximally
for intermediate values of the activity.
In any case, the above considerations prove that no mapping exists between
the correlation functions of the disordered system at finite activity
and those of the pure system at the same density, $\rho(\rs,\lambda)$. 


\section{Approximate schemes}

The  lattice model   studied here is exceptional   in  that we know its exact
solution, even when disorder is present.  More  often, we will have only a
partial description, e.g. the  cluster expansion  that applies at  low
site  density in the case of  the RSS \cite{OT07}.  In these cases, we
seek approximate schemes that can  provide an accurate description of
the  adsorption isotherms  for the  full  range of site  densities and
activities. In this section we propose two such schemes and evaluate their
accuracy by comparing with the exact results.  

One approach involves a partial resummation of 
the site density expansion  of Eq.(\ref{eq:11}). Specifically, 
retaining explicitly the first term that appears at all orders we can write
\begin{align}\label{eq:17}
\rho(\rho_s,\lambda)&=\sum_{l=1}^{\infty}\left(-
\left(\frac{-\lambda}{1+\lambda}\right)^l +F_l(\rs,\lambda)\right)\rho_s^l
\end{align}
where $ F_l(\rs,\lambda)$ represents the remaining terms in the exact expansion and,
consistent with these terms, 
has the property that
\begin{equation}
F_l(\rs,\lambda)\to 0
\end{equation}
when $\lambda\to 0 $ and $\lambda\to \infty$.

\begin{figure}

\resizebox{8cm}{!}{\includegraphics{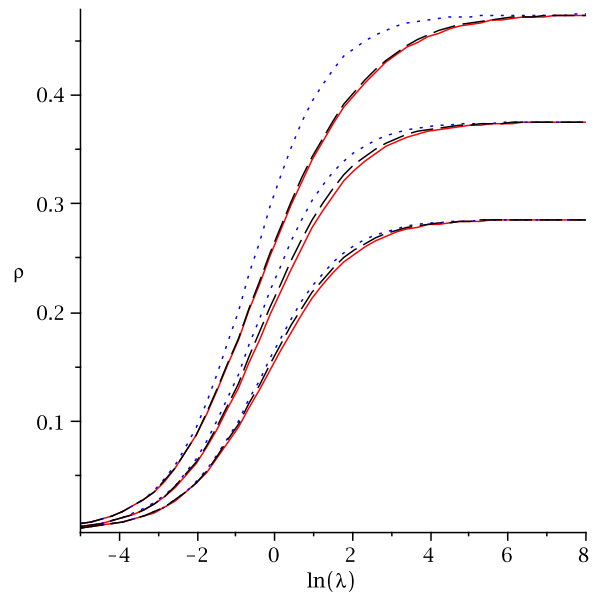}}
\caption{Density $\rho$ as a function of activity $\lambda$ for $\rho_s=0.4,0.6,0.9$
calculated via the exact formula, Eq.(\ref{eq:11}) (full curves), via Eq.(\ref{eq:12}) (dotted lines), and via Eq.(\ref{eq:18}) to the zeroth order ($f(\rs,\lm)=1$) (dashed curve) (see text).}\label{fig:rho}
\end{figure}

The simplest approximation is to set 
$F_l(\lambda)=0$, resulting in the approximate isotherm
\begin{equation}\label{eq:12}
\rho(\rho_s,\lambda)={\frac {\lambda\,{ \rho_s}}{\lambda\,{ \rho_s}+1+\lambda}}
\end{equation}
that, as  expected,
has the correct behavior in the limits of small    and large
activities. 

In order  to  highlight the deviations from the  exact  results, we have
plotted in Fig.\ref{fig:rho} the density  $\rho$ as a function of  the
activity $\lambda$  for    various   values  of the   site     density
$\rho_s$. The approximate isotherms  always overestimate  the adsorbed
density  for all activities.  The  deviations  increase with the  site
density and are most pronounced  at intermediate activities. In
order to understand the origin of this  discrepancy, we have performed
an asymptotic expansion of Eq.(\ref{eq:11}) at large activity,
\begin{equation}\label{eq:14}
\rho(\rho_s,\lambda)\simeq\frac{\rho_s}{1+\rho_s}-\frac{1}{6\lambda}\frac{\rho_s(-\rho_s^3+3\rho_s+6)}{(1+\rho_s)^3(1-\rho_s)}+...
\end{equation}
and compared with that of Eq. \ref{eq:12},
\begin{equation}\label{eq:15}
\rho(\rho_s,\lambda)\simeq\frac{\rho_s}{1+\rho_s}-\frac{\rho_s}{\lambda(1+\rs)^2}+...
\end{equation}
We note that  when $\rho_s\to 0$, Eq.(\ref{eq:14}) and Eq.(\ref{eq:15}) coincide. 
  This
explains the very accurate description of the approximate isotherm in this
limit. Conversely, when $\rho_s \to 1$, one obtains
\begin{equation}\label{eq:16}
\rho(\rho_s,\lambda)\simeq\frac{\rho_s}{1+\rho_s}-\frac{1}{6(1-\rho_s)\lambda},
\end{equation}
whereas no such combination in $\lambda(1-\rs)$ appears in Eq. (\ref{eq:15}).

By including   the     leading term   of   the  asymptotic   behavior,
Eq.(\ref{eq:14}), the agreement with the exact  result is only correct
up to intermediate  $\rho_s$.  It would be   necessary to add additional
terms when $\rho_s$ goes to $1$ in order to  have a good approximation of
the isotherms. Such a  procedure rapidly becomes very  complicated and is
not useful for more realistic models. The asymptotic analysis seems to
indicate that the   neglected  terms of   the series  correspond to   the
existence of an asymptotic series whose variable is $((1-\rho_s)\lambda)^{-1}$.

\begin{figure}[t]
\resizebox{8cm}{!}{\includegraphics{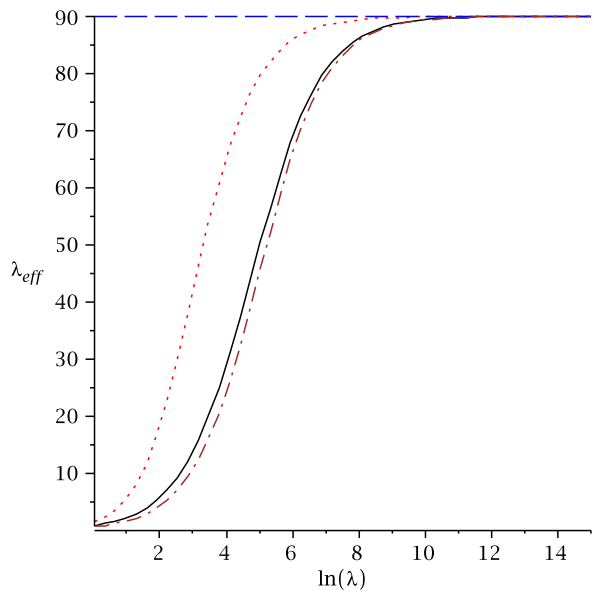}}
\caption{Effective activity $\lmeff$ as a function of 
$\lambda$ for $\rho_s=0.9$ calculated via the exact formula, Eq.(\ref{eq:11}) (full curve),
 via Eq.(\ref{eq:18}) to the zeroth order (dotted curve) and to the first-order, Eq.(\ref{eq:19}) (dotted-dashed curve).}\label{fig:7}
\end{figure}

An alternative approach uses an effective activity, 
$\lmeff$,   instead  of  $\rho(\rho_s,\lambda)$. The idea is to
estimate the density in the disordered system with
\begin{equation}\label{eq:leff}
\rho(\rs,\lm)=\rho^*(\lmeff(\lm,\rs))
\end{equation}
where   $\rho^*$  corresponds to  the  density  at equilibrium given by
Eq.(\ref{eq:13}). Note that, since there is no exact mapping to a pure system for a finite activity $\lambda$, 
even if we could find the function $\lmeff(\lm,\rs)$
that satisfies Eq. \ref{eq:leff}, it would not give exact results 
for other thermodynamic quantities. 
There is, however, merit in this approach because, as we
have shown  in previous sections, there is a mapping 
in the limits of large  and
small  activities.
Thus,  by  using
Eq.(\ref{eq:14}) with Eq. (\ref{eq:13}), one obtains the asymptotic expansion
\begin{equation}
\frac{1}{\lmeff}={\frac { \left( -1+\rho_s \right) ^{2}}{\rho_s}}+{\frac {6+3\,
\rho_s-{\rho_s}^{3}}{6\rho_s\lambda}}+O \left( {\lambda}^{-2} \right) 
\end{equation}
The coefficient of the leading term of this expansion is
a function   of  $\rho_s$ which behaves  simply   in  the two limits
$\rho_s\to 0$ and  $\rho_s\to 1$. By combining  with
the    low-activity  expansion ($1/\lmeff=1/(\lambda\rho_s)$),  we
propose the following interpolation scheme
\begin{equation}\label{eq:18}
\frac{1}{\lmeff}={\frac { \left( -1+\rho_s \right) ^{2}}{\rho_s}}+\frac{f(\lambda,\rho_s)}{\lambda \rho_s}
\end{equation}
where $f(\lambda,\rho_s)$ is a $[n,n]$ Pad\'e  approximant. The isotherms are then calculated
using Eq. (\ref{eq:leff}). As can be seen in Fig. \ref{fig:rho},
even to  zeroth order  ($f(\lambda,\rho_s)=1$), this route gives a 
significantly better approximation  than Eq.(\ref{eq:12}). 
By matching  the exact asymptotic  behavior, Eq.(\ref{eq:16}),
as well as the second-order low-activity expansion, one obtains 
\begin{equation}\label{eq:19}
f(\lambda,\rho_s)=\frac{
  1+a(\rho_s)
\lambda }{   1+b(\rho_s)\lambda}
\end{equation}
where
\begin{equation}
a(\rho_s)=\frac {6+ 2\,{\rs}^{3}+9\,{\rs}^{2}-{\rs}^{5
}+9\,\rs-{\rs}^{4}  }{6-3\,\rs-{\rs}^{3}}
\end{equation}
and 
\begin{equation}
b(\rho_s)=6\,\frac {\rs\, \left( 1+\rs+{\rs}^{2}
 \right)}{6-3\,\rs-{\rs}^{3}}.
\end{equation}

Since  the zeroth order approximation is practically indistinguishable   from the exact
result in the isotherm plot (Fig. \ref{fig:rho}), we highlight the differences between the two 
by plotting  the effective activity versus the
activity for high density $\rho_s$ in Fig. \ref{fig:7}. The horizontal
line corresponds  to the asymptotic value,  the full  line to the exact
result,  the dotted line  to the zeroth order, and  the dashed line to
the first-order.  We note that  this method converges very rapidly  to
the exact result, even at high $\rho_s$, which is very encouraging for
systems where no exact result is known.

We can thus conclude from the present study that (i) There is a mapping
between the adsorbed configurations of dimers on a diluted 
one-dimensional lattice and those on a full lattice at the same adsorbed
density in the two limits of small and large activities; (ii) This
mapping does not extend to finite activities. However, a successful approximation
scheme is provided by considering a pseudo-mapping through the introduction of
an effective activity. The scheme has been tested on the present 
one-dimensional lattice system for which we have also obtained exact expressions, 
and work is in progress to apply it to more realistic situations.

\appendix
\section{Pure model}

\begin{figure}[h]

\resizebox{6cm}{!}{\includegraphics{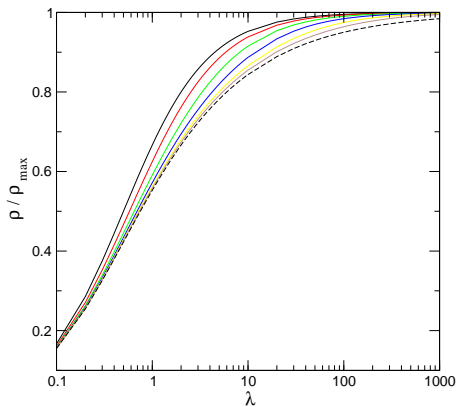}}
\caption{Adsorption isotherms on a pure chain of N sites ($N=2,4,8,16,32,64$ top to bottom). 
The dashed line is the thermodynamic limit.}\label{fig:isotherms}
\end{figure}

To obtain  an analytic  expression for the  partition function  of the
pure    system   we  use     the   transfer   matrix    method:   see,
e.g. \cite{G89}. Using this approach, Eq.(\ref{eq:1}) can be rewritten
as
\begin{equation}
\Xi^*(\lambda,N)=Tr(T^NA)
\end{equation}
where
\begin{align}
T=\left(\begin{array}{cc}
	1&1\\
	\lambda&0
	\end{array}
\right)
\end{align}
and 
\begin{align}
A=\left(\begin{array}{cc}
	0&1\\
	1&1
	\end{array}
\right).
\end{align}
Performing the trace operation, the partition function becomes
\begin{equation}
\Xi^*(\lambda,N)=\frac{\lambda_1^{N+1}-\lambda_2^{N+1}}{\lambda_1-\lambda_2}+\lambda\frac{\lambda_1^N-\lambda_2^N}{\lambda_1-\lambda_2}
\end{equation}
where $\lambda_1$ and $\lambda_2$ are the eigenvalues of the matrix $T$,
\begin{equation}
\lambda_{1,2}=\frac{1\pm\sqrt{1+4\lambda}}{2}.
\end{equation}
For finite  $N$ the partition function is  a polynomial in $\lambda$, e.g.
$\Xi^*(\lambda,1)=1+\lambda$, $\Xi^*(\lambda,2)=1+2\lambda$, $\Xi^*(\lambda,3)=1+3\lambda+\lambda^2$, $\Xi^*(\lambda,4)=1+4\lambda+3\lambda^2$,...
In the thermodynamic limit, only the largest eigenvalue contributes to
the thermodynamic quantities and we have
\begin{equation}
\lim_{N\to\infty} \frac{\ln(\Xi^*(\lambda,N))}{N}=\ln\left(\frac{1+\sqrt{1+4\lambda}}{2}\right).
\end{equation}

Some isotherms for different values of $N$ are shown in
Fig. \ref{fig:isotherms}. Note that the thermodynamic limit is
approached rather slowly.


\end{document}